\documentclass[pdflatex,sn-basic]{sn-jnl}% Basic Springer Nature Reference Style/Chemistry Reference Style

\usepackage{lineno}
\usepackage{hyperref}
\usepackage{soul}
\usepackage{tabularx}
\usepackage{array}
\usepackage{epsfig}
\usepackage{ulem}
\usepackage{xcolor}
\usepackage{tikz}
\usepackage{pgf-pie}
\usepackage{caption}
\usetikzlibrary{shapes.geometric,patterns}
\usepackage{pgfplots}
\pgfplotsset{compat=1.17} % 使用最新的兼容性设置
\usepackage{booktabs}
\usepackage{multirow}
\usepackage{amssymb}
\usepackage{amsmath}
\usepackage{amsthm}
\usepackage{graphicx}
\usepackage{rotating} % 引入旋转宏包
\usepackage{caption}
\usepackage{booktabs} % 推荐，使表格看起来更专业
\usepackage{longtable} % 如果表格需要跨页，可以使用 longtable
\usepackage{multirow}
\usepgfplotslibrary{colorbrewer}
\usepackage{graphicx}%
\usepackage{multirow}%
\usepackage{amsmath,amssymb,amsfonts}%
\usepackage{amsthm}%
\usepackage{mathrsfs}%
\usepackage[title]{appendix}%
\usepackage{xcolor}%
\usepackage{textcomp}%
\usepackage{manyfoot}%
\usepackage{booktabs}%
\usepackage{algorithm}%
\usepackage{algorithmicx}%
\usepackage{algpseudocode}%
\usepackage{listings}%
\usepackage{natbib}
\theoremstyle{thmstyleone}%
\theoremstyle{thmstyletwo}%

\theoremstyle{thmstylethree}%

\begin{document}

\title[Article Title]{Unlocking Learning Potentials: The Transformative Effect of Generative AI in Education Across Grade Levels}

\author[1]{\fnm{Meijuan} \sur{Xie}}

\author*[1]{\fnm{Liling} \sur{Luo}}\email{593929431@qq.com}

% \affil*[1]{\orgdiv{School of Mathematics and Statistics}, \orgname{Guangxi Normal University}, \orgaddress{\street{ Yanshan Street}, \city{Guilin}, \postcode{541004}, \state{Guangxi Zhuang Autonomous Region}, \country{China}}}

\affil*[1]{\orgdiv{School of Mathematics and Statistics}, \orgname{Guangxi Normal University}, \orgaddress{\city{Guilin}, \postcode{541004}, \country{China}}}

\abstract{The advent of generative artificial intelligence (GAI) has brought about a notable surge in the field of education. The use of GAI to support learning is becoming increasingly prevalent among students. However, the manner and extent of its utilisation vary considerably from one individual to another.  And researches about student's utilisation and perceptions of GAI remains relatively scarce. To gain insight into the issue, this paper proposed a hybrid-survey method to examine the impact of GAI on students across four different grades in six key areas (LIPSAL): learning interest, independent learning, problem solving, self-confidence, appropriate use, and learning enjoyment. Firstly, through questionnaire, we found that among LIPSAL, GAI has the greatest impact on the concept of appropriate use,  the lowest level of learning interest and self-confidence. Secondly, a comparison of four grades revealed that the high and low factors of LIPSAL exhibited grade-related variation, and college students exhibited a higher level than high school students across LIPSAL. Thirdly, through interview, the students demonstrated a comprehensive understanding of the application of GAI. We found that students have a positive attitude towards GAI and are very willing to use it, which is why GAI has grown so rapidly in popularity. They also told us prospects and challenges in using GAI. In the future, as GAI matures technologically, it will have an greater impact on students. These findings may help better understand usage by different students and inform future research in digital education.}

% RQ1GAI对合理使用影响最大、学习兴趣和自信心最低
% RQ2在不同年级中，LIPSAL的影响因素高低不相同，高年级普遍高于低年级
% RQ3


\maketitle

\section{Introduction}\label{sec1}

In the context of the accelerated innovation in technology, Generative artificial intelligence (GAI) is becoming a significant driver of social advancement \citep{lv2023generative}.  GAI represents a novel approach to artificial intelligence (AI), whereby machine learning models are employed to generate new data, content, pictures and even videos. Specifically GAI is able to simulate the distributional characteristics of data by learning large amounts of data to create entirely new content that is similar to the training data, but not identical \citep{jovanovic2022meaning}. This technology is different from previous technologies in that it is innovative and versatile, so it can play a huge role in a variety of areas such as artistic creation, data processing and personalised services \citep{banh2023ways, teubner2023welcome}. For instance, in film production, GAI can generate realistic special effects and scenes for an immersive viewing experience. However, the development of GAI also brings a series of challenges, such as ethics and copyright issues \citep{Luke2023danger}. When utilising the capabilities of GAI, it is essential to use it in a prudent and judicious manner, seeking to enhance its strengths while avoiding its weaknesses.

ChatGPT is an artificial intelligence chatbot developed by OpenAI. It is based on advanced deep learning technology and has the ability to have natural conversations with humans \citep{Dempere2023openAI,wu2023openai, Wardat2023math}. Natural Conversation represents ChatGPT's capacity to emulate human discourse and engage in seamless communication with users, offering services such as information retrieval and question-answering. During the dialogue, it also constantly adjusts its answers  based on the user's feedback in order to better meet the user's needs. Due to its extensive knowledge acquired through extensive textual data training, it is able to comprehend and generate expressions in multiple languages, thereby exhibiting a high degree of intelligence in communication and interaction \citep{Roumeliotis2023text}. As a model based on deep learning, ChatGPT represents the latest trend in natural language processing technology and provides a new direction for research in the field of AI \citep{Lin2023trend}.

\cite{wu2023openai} noted a significant surge in the utilisation of GAI, accompanied by a multitude of applications that have attracted worldwide attention. The continued integration of GAI with the education industry has shown some positive impact, with an increasing number of students opting to utilise it as a learning assistant \citep{wong2020aieducation,Gocen2020Education}. Everything has two aspects: a positive impact and a negative impact. So this article focuses on the use of GAI (such as ChatGPT \footnote{\url{https://chatgpt.com/}}, Kimi \footnote{\url{https://kimi.moonshot.cn/}}, Doubao \footnote{\url{https://www.doubao.com/}}) by students to gain insight into their comprehension of GAI and to facilitate more effective assistance in their learning. The LIPSAL is to evaluate the effectiveness of GAI in assisting students in their learning and overall development, which are extremely important in learning with GAI \citep{Liu2022interest, Cheng2023confidence}. They have the potential to facilitate students' holistic development to a significant extent. Therefore, the article proceeds to investigate the differences in the extent to which the GAI improves LIPSAL for students at different grade levels. In order to fulfill the requirements of these studies, the paper was based on three key research questions: 

\begin{itemize}
\item \textbf{RQ1:} What extent does GAI impact students in LIPSAL ?
\item \textbf{RQ2:} What are the differences of LIPSAL between four grade levels in using GAI ?
\item \textbf{RQ3:} What are the perceptions and understanding of students regarding GAI ?
\end{itemize}

\section{Literature Review}\label{sec2}

\subsection{The application of GAI in education}
With the continuous development of artificial intelligence (AI) technology, researches on how to use ChatGPT in the field of education continues to deepen. The use of AI in education to tailor learning for each student can replace the one-size-fits-all approach. Meanwhile, the advancement of education can also promote the iteration and revolution of AI and related supporting tools. AI and these tools will be of great help in assisting teachers and students. For educators, \cite{Emma2024app} point out that teachers choose appropriate and high quality educational applications when teaching, to improve quality of teaching and learning. \cite{meyer2023chatgpt} assert that AI has the potential to transform the pedagogical approaches and educational assessment, to facilitate the acquisition of feedback on their teaching and to enable them to refine their methods. For students, \cite{Ali2024agency} identify a growing utilisation of AI-powered learning technologies. These technologies are employed in a variety of ways, including the construction of personalised learning pathways, the provision of real-time feedback for asked questions, and the formulation of recommendations regarding optimal study time and content. The evidence suggests that these technologies have a beneficial impact on student learning. Furthermore, AI can be used in different fields, such as improving mathematical thinking skills \citep{frieder2023mathematical}, 
 assisting in language learning \citep{Jaeho2023english}, and writing instruction \citep{Mei2024writing}. \cite{zhang2023Higher_Education} introduced ChatGPT's advantages and functions in detail, he also pointed out some challenges and concerns in future development, as well as its impact on educational work. In the field of education, research on GAI is still in its infancy, as there are few empirical studies investigating effective learning framework or the application of learning strategies in using GAI \citep{Chung2024review}. So how to use GAI in education appropriately still requires further in-depth research, which will show a positive trend for the future development of education.

\subsection{GAI's improvement on students' learning in LIPSAL}
The LIPSAL is of significant importance with regard to the student learning process. The following LIPSAL was to be presented, as indicated in Table \ref{tab:six aspects}. \cite{Liu2022interest} showed that AI was able to mimic human beings in communication and dialogue, and has a high degree of social ability. It can give students timely and positive feedback during dialogue, improving the interactive experience. Thus, it makes students keep a strong willingness to communicate with it, maintaining students' learning interest. AI-powered systems will help content keep up with learners, help systems better identify learner needs, end waste of time and resources, enable rapid data analysis, and improve students’ problem-solving abilities \citep{karamustafaouglu2023developing}. \cite{chassignol2018artificial} mentioned that using AI-powered predictive computing can understand students’ habits and propose the most effective study schedule for them. These personalized designed schedules will guide students to mastery and review lessons as needed as a way to increase their ability to learn independently. Students that have developed an interest in learning with the help of GAI, or who have been rewarded for their achievements, have increased levels of both internal and external motivation, which will significantly enhance their self-confidence in learning \citep{egara2024effect}. \cite{Wardat2023math} observed that the prevailing view of GAI in  educational is positive. The study revealed that the utilisation of GAI by a significant proportion of the participants led to an enhanced learning experience and the potential to enhance students' abilities, thereby increasing their motivation to learn. They also emphasised that GAI is a valuable educational instrument, but that it should be employed with caution and that guidelines for its prudent use should be established.

 \vspace*{-0.08cm}
\begin{table*}[!t]
 \caption{Definition of the six aspects of LIPSAL.}
 \label{tab:six aspects}
    \centering
    \small
\scalebox{.95}
    {    
    \setlength{\extrarowheight}{2.8pt}
    \begin{tabular}{p{4.5cm}p{8cm}} % 保持列宽不变，LaTeX 会自动处理换行
    \toprule
    \textbf{LIPSAL}
     & Descriptions \\
    \midrule
    \text{\textbf{Learning interest}} & The level of interest a student demonstrates in the learning process is indicative of their tendency of the act of learning. The cultivation of interest in learning is a crucial aspect of the educational process \citep{Cheng2023confidence}.\\
    \text{\textbf{Independent learning}} & It can be defined as self-directed and effective learning. It is essential to provide students with access to independent learning resources and environments\citep{lin2024effects}.\\
    \text{\textbf{Problem solving}} & It is used to describe the utilisation of methodologies aimed at resolving issues. It is frequently associated with the capacity to innovate \citep{Gocen2020Education}. \\
    \text{\textbf{Self-confidence}} & A self-confident individual possesses a belief and trust in their own capacity to learn, and this contributes to a higher level of motivation \citep{almasri2022simulations}. \\
    \text{\textbf{Appropriate use}} & It is imperative to adhere to the pertinent legislation, and ethical standards, and to employ them in an appropriate and effective manner \citep{rodriguez2024artificial}. \\
    \text{\textbf{Learning enjoyment}} & It can be defined as the positive emotions and feelings experienced by students during the learning process. These feelings contribute to an enhanced willingness to learn among students \citep{pham2024learning}. \\
    \bottomrule
    \end{tabular}
}
    \vspace*{0.1cm}
    \vspace*{-0.1cm}
\end{table*}

\subsection{Comparison of the LIPSAL of learning across different grades} 
Students of different ages learn at different developing rates \citep{wong2022age}. There are still relatively few comparative studies between grades, but some attempts are being made in this area. \cite{Liou2021grade_interest} studied the same sample of students in 2011 and 2015 respectively. Their results showed that these students in primary school had a stronger interest and enjoyment in learning than they did in junior high school, also have greater self-confidence. However, interest is very important and students who have a strong interest in learning will tend to exhibit greater academic growth \citep{schubatzky2023predicting}. So as the grade level increases, they will be less interested in learning and feel more negative about their educational experience \citep{Ding2007negative}, which can have a detrimental impact on their academic development. \cite{teppo2021value} observed that while students in the upper grades may demonstrate a diminished interest, they will focus more on the value and practical applicability of subject knowledge, facilitating the transformation of intrinsic motivation into extrinsic motivation. \cite{zimmerman1990use} conducted a comparative analysis of students at the elementary, middle, and high school. The study found that students in higher grades exhibited significantly greater problem-solving abilities and independence in learning than their counterparts in lower grades. Furthermore, they are also more proficient in the utilisation of assisted learning applications, exhibiting enhanced capabilities in self-regulation. However, they are less purpose orientated in their endeavours and less motivated to learn \cite{Tang2008orientation}. So while teaching students knowledge, it is more important to focus on the development of interest in learning.

While there is a substantial body of literature examining the impact of GAI on students, there are several shortcomings in this existing research. Some studies lack insight into students' perceptions of GAI, while others fail to compare the use of GAI across different student groups. Additionally, many studies focus on a single influencing factor, limiting the scope of their analysis. This paper therefore builds on existing studies with in-depth analyses to investigate the impact that different students' learning with GAI has on LIPSAL, and also examining students' perceptions regarding this learning experience.

% \section{Conceptualising the case study}\label{sec3}

\section{Methodology}\label{sec4}

The purpose of this study is to investigate students employ GAI as a tool for learning about the impact on the LIPSAL and for comparing the level of impact on the LIPSAL across grades. Additionally, it is used to ascertain students' perceptions of their opinions. The study uses a primarily questionnaire-based approach, complementing by group interviews to gain a more in-depth understanding of the study's validity and reliability. Therefore, the study is divided into two stages: questionnaire and interview content analysis.

\begin{itemize}
\item \textbf{Dimension 1}: Questionnaire analysis

\item \textbf{Dimension 2}: Interview content analysis
\end{itemize}

\begin{figure}[t]%% placement specifier
\centering%% For centre alignment of image.
\includegraphics[width=0.99\textwidth]{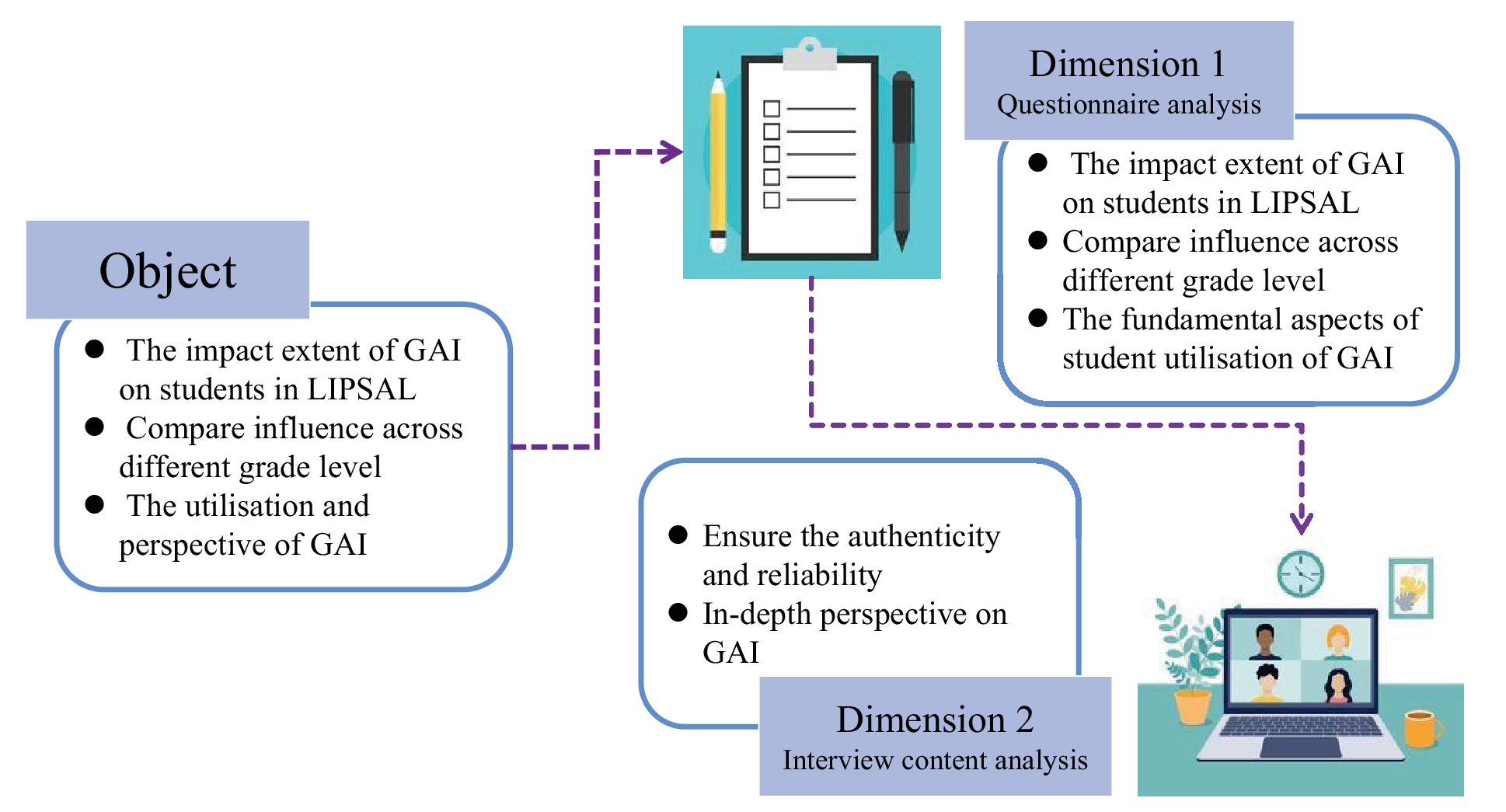}
\caption{Structure of the proposed hybrid-survey research method.}
\label{framework}
\end{figure}

Fig.\ref{framework} illustrates the concept of the study, demonstrating the link between the two aspects under investigation. The first area focuses on the fundamental aspects of student utilisation of GAI, whether the GAI effectively promotes the LIPSAL and differences between students at different grade levels. The second area of investigation concerns students' perceptions of the content analysis of the interviews, providing insights into their understanding and experiences of learning the GAI.

\subsection{Research setting}
\label{3.1}

This experiment is based on different grades of students, including Sophomores in high school,juniors in high school, juniors and seniors in university, a total of 90 people. Firstly, the experimental questionnaire is developed based on the feedback from eight in-service teachers who use GAI to teach and two computer professionals to ensure the validity of the experimental survey. After evaluating the reasonableness of the questionnaire, the survey research is implemented. The experiment will be implemented through an online survey conducted with the Wenjuanxing platform \footnote{\url{https://www.wjx.cn/}} from July to August 2024. Secondly, in order to ensure the authenticity and reliability of the results, group interviews are organized to record an in-depth understanding of student's specific use of GAI, including the impact they have made in using it and their views on its development. The final instrument used to collect data is a five-point Likert scale questionnaire \citep{likert1932fivepoint}. It ranges from “definitely disagree” scores one point to “definitely agree” scores five points. An open-ended qualitative question is included in the survey to further investigate student attitudes, experiences,and perceptions. Inspired by \cite{Ali2023Motivation}, in another section, consists of fifteen declarative sentences, attempts to discover whether GAI can increase students' learning interest, independent learning, problem solving, confidence, appropriate use, and learning enjoyment.

\subsection{Participants}
\label{3.2}

A total of 90 volunteers participated in the study. In order to guarantee the usability of the questionnaire, volunteers were selected who had utilised GAI and possessed a certain depth of knowledge regarding its functionality. We took into account that high school and university students use GAI more extensively and in greater depth than primary school students. Additionally, in order to control the variables of the study, we selected students from a Beijing high school in their sophomore and junior years, as well as students from a university in their junior and senior years. As shown in Table \ref{tab: participant} , for grades, the sample comprised 15 sophomores in high school (16.67\%), 15 juniors in high school (16.67\%), 30  juniors in university (33.33\%), and 30 seniors in university (33.33\%). As to genders, it was composed of 50\% males and 50\% females. Subsequently, 20 students were randomly selected for participation in group interviews. In order to eliminate potential external influence, students were not apprised of the subject matter in advance, they were not permitted to utilize electronic devices to respond to the queries, and each student was individually interviewed.

\vspace*{-0.08cm}
\begin{table*}[!h]
 \caption{The gender and grade distribution of participants (N=the number of students).}
 \label{tab: participant}
    \centering
    \small
\scalebox{1.0}
    {    
    \setlength{\extrarowheight}{3pt}
    \begin{tabular}{l@{\hspace{1.6cm}}c@{\hspace{1.6cm}}c@{\hspace{1.6cm}}c@{\hspace{1.6cm}}}
    \toprule
    \textbf{Category} & \text{Items} & N & Proportion(\%) \\
     \midrule
    \text{\textbf{Gender}} & \text{\text{Male}}&  45 & 50 \\
     & \text{\text{Female}} & 45 & 50  \\
     % \midrule
    \text{\textbf{Grade}} &
    \text{\text{Sophomores in H.S}} & 15 & 16.67\\
     & \text{\text{Juniors in H.S}} & 15 & 16.67\\
     & \text{\text{Juniors in U.S}} & 30 & 33.33 \\
     & \text{\text{Seniors in U.S}} & 30 & 33.33\\
    \bottomrule
    % \vspace*{0.05cm}
    \end{tabular}
}

\vspace{0.5em}
\centering 
\footnotesize
\text{Note: H.S represents High School and U.S represents University.} 
    % \vspace*{0.1cm}
    % \vspace*{-0.1cm}
\end{table*}

\subsection{Data collection and analysis}
\label{3.3}

There are two ways to collect data. Firstly, since the participants are not in the same area, data are collected through an online questionnaire through Wenjuanxing platform. It is an online questionnaire survey and data collection tool that provides a convenient way to design questionnaires, collect data, and conduct preliminary data analysis. The questionnaire consisted a total of 18 questions, the initial three of which sought information about the volunteer's gender, age, and frequency of GAI utilisation. The following 17 questions investigate the LIPSAL, each comprising two or three questions. The LIPSAL represent the six aspects of this study. Question 18 was to elicit students' views on a given topic through an open-ended inquiry. Secondly, use Tencent Meeting \footnote{\url{https://meeting.tencent.com/}} to conduct group interviews and communicate in depth with students. It is an online video conferencing software suitable for various scenarios such as remote communication, meetings, teaching and group interviews. The interview questions were four open-ended inquiries designed to elicit insight into the positive and negative effects of students' use of GAI, their attitudes towards it, and their perceptions of its future development.

The collected data are analyzed quantitatively and qualitatively. During quantitative analysis, we set important variables for each question, and view the distribution of those variables with data entry. Then we apply SPSS 26.0 for descriptive statistical to test the overall distribution of variables and a analysis of variance (ANOVA) to comparing differences in variables, in order to find out the impact in the LIPSAL from GAI, and the differences between grades. Moreover,we conducted a Pearson’s correlation analysis to further understand the relationship between internal aspects of LIPSAL when utilising
GAI. During the qualitative analysis, word frequency analysis is performed on the open-ended questions to identify the main themes and sub-themes to gain an in-depth understanding of students’ attitudes towards GAI learning.

\section{Results and Discussion}\label{sec5}

\subsection{RQ1: What extent does GAI impact students in the LIPSAL?}
\label{4.1}
This paper analyses the extent to the LIPSAL when 90 students utilise GAI to facilitate their learning. As shown in Table.\ref{tab: attitude}, each variable contains 2 or 3 questions. Each student provided a response to each variable in accordance with their particular circumstances. In general, for the impact of GAI's assistance in the LIPSAL, the number of respondents who expressed agreement was significantly higher than the number of those who expressed disagreement or neutrality.

\vspace*{-0.08cm}
\begin{table*}[!h]
 \caption{Proportion of students with different attitudes in relation to LIPSAL.}
 \label{tab: attitude}
    \centering
    \small
\scalebox{1.0}
    {    
    \setlength{\extrarowheight}{3pt}
    \begin{tabular}{l@{\hspace{.6cm}}l@{\hspace{.3cm}}c@{\hspace{.3cm}}c@{\hspace{.3cm}}c}
    \toprule
    \textbf{LIPSAL}
     & Paired questions & Disagree(\%) & Neutral(\%) & Agree(\%)
     \\
     \midrule
    \text{\textbf{learning interest}} & Q 4 \& 5 & 1 & 17 & 82  \\
    \text{\textbf{independent learning}} & Q 6 \& 7 & 3 & 20 & 77  \\
    \text{\textbf{problem solving}} & Q 8 to 9 & 2 & 19 & 79 \\
    \text{\textbf{self-confidence}} & Q 10 \& 11 & 3 & 22 & 75  \\
    \text{\textbf{appropriate use}} & Q 12 to 14 & 2 & 11 & 87  \\
    \text{\textbf{learning enjoyment}} & Q 15 to 17 & 1 & 19 & 80  \\
    \bottomrule
    \end{tabular}
}
    \vspace*{0.1cm}
    \vspace*{-0.1cm}
\end{table*}

\begin{figure}[t]%% placement specifier
\begin{tikzpicture}
\begin{axis}[
    xbar,
    enlargelimits=0.15,
    width=10cm, % 设置图表宽度
    height=6cm,
    legend style={at={(0.5,-0.15)},anchor=north,legend columns=-1},
    symbolic y coords={Self-confidence, Learning interest, Problem solving, Learning enjoyment, Independent learning, Appropriate use},
    xlabel={Mean value (M)}, 
    xmin=3.9, % 设置横坐标的最小值
    xmax=4.1, % 设置横坐标的最大值
    xtick={3.9, 4.0, 4.1}, % 设置横坐标的刻度
    nodes near coords,
    nodes near coords align={horizontal},
    ytick=data,
    ]
\addplot[pattern=dots] coordinates {(3.96,Self-confidence) (3.96,Learning interest) (4.0,Problem solving) (4.01,Learning enjoyment) (4.08,Independent learning) (4.09,Appropriate use)};
\end{axis}
\end{tikzpicture}
\caption{The mean value of LIPSAL of the questionnaire.}
\label{affect}
\end{figure}
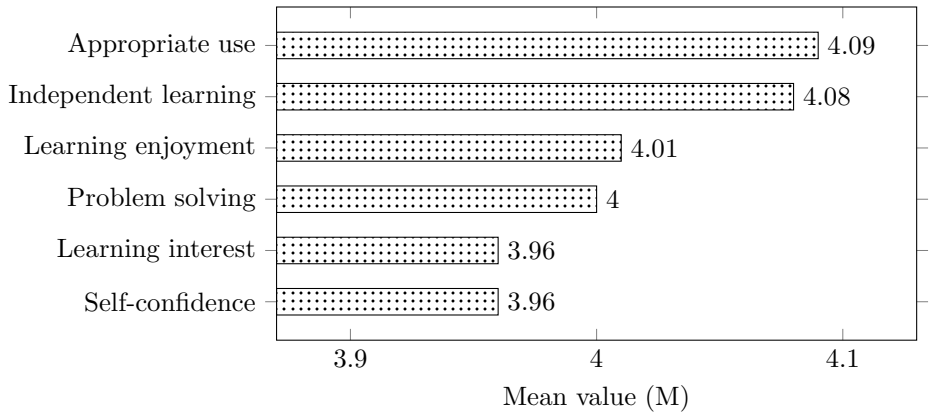

We calculated the corresponding mean values of the LIPSAL from the questionnaire responses. According to the principle of questionnaire scoring, we agreed that GAI has a positive facilitating effect on the corresponding variable aspect when the mean value is greater than 3 (M \textgreater 3), and called weak facilitation (3 \textless{} M \textless{} 4) and strong facilitation (M $\geq$ 4). The greater the value, the better the facilitating effect. As shown in Fig.\ref{affect}, the M value of all the LIPSAL are higher than 3, which indicates that it has a facilitating effect on all of them. But there are still some differences between them, the higher mean values were observed for appropriate use (M=4.09) and independent learning (M=4.08). They indicate that 1) most students are carefully monitored and that instances of plagiarism and other forms of academic dishonesty are rare, 2) a notable part of students are afforded the opportunity to engage in self-directed learning planning and to monitor their progress through the utilisation of GAI. However, the lowest mean values were observed for interest in learning (M=3.96) and self-confidence (M=3.96). It indicate that students' interest in the subject matter has not been significantly enhanced, nor have they experienced an increase in self-confidence in the process of learning. Therefore, when integrating GAI with education, it is imperative to prioritise the enhancement of students' motivation to learn. This is an area that warrants further investigation and improvement.

\begin{figure}[ht!]%% placement specifier
\centering%% For centre alignment of image.
\includegraphics[width=0.98\textwidth]{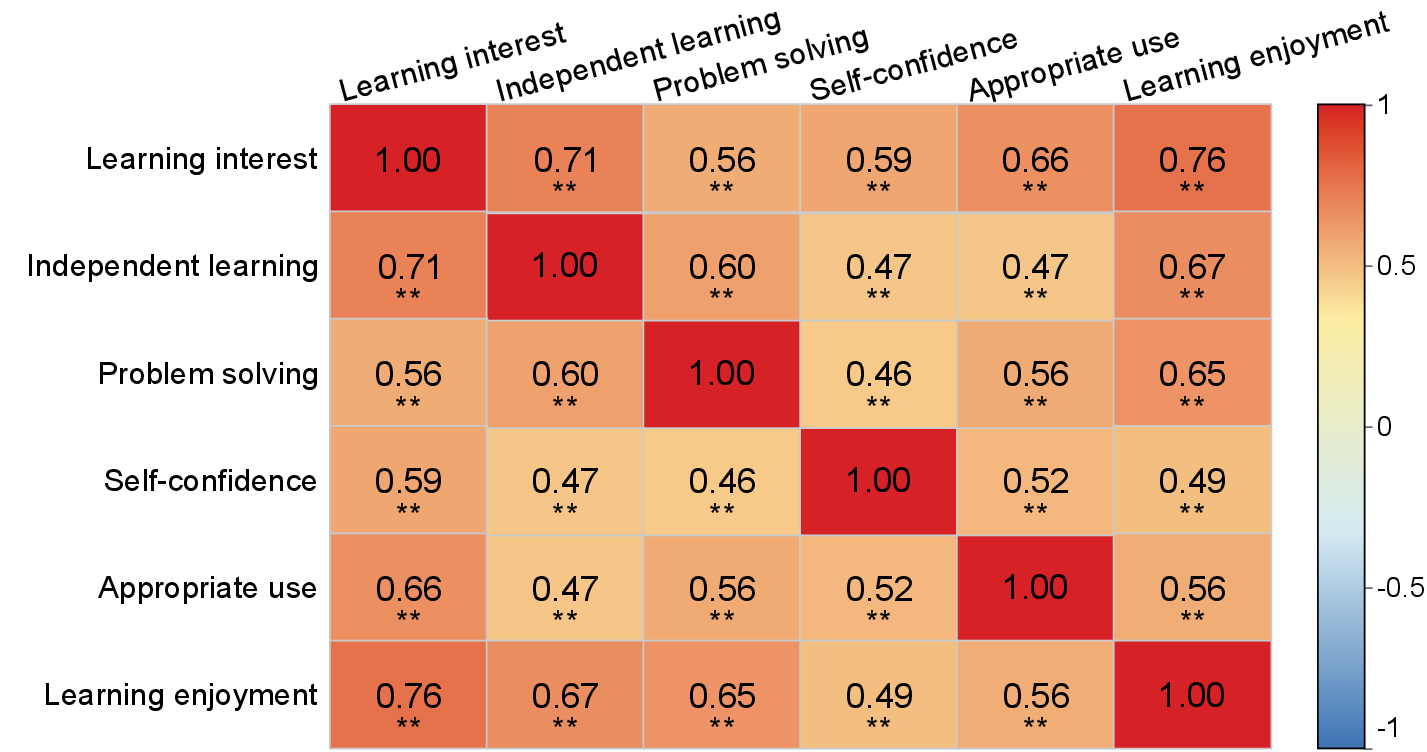}
\caption{Correlation between internal aspects of LIPSAL.}
\label{correlation}
\centering 
\footnotesize
\text{Note: ** represents p \textless{} 0.01.} 
\end{figure}

\begin{figure}[ht!]%% placement specifier
\centering%% For centre alignment of image.
\includegraphics[width=1\textwidth]{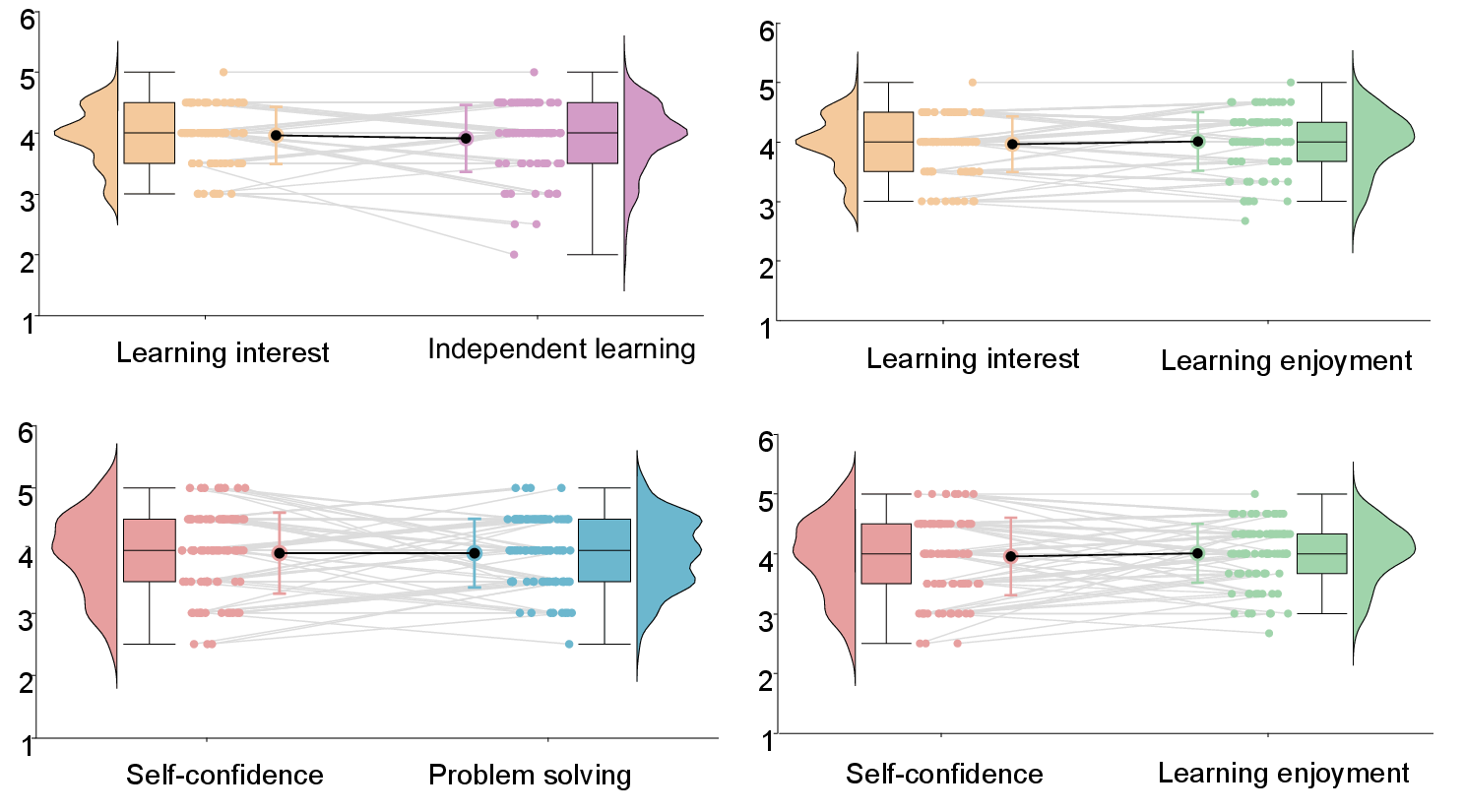}
\caption{Comparison of the distribution of scores between two variables.}
\label{detail}
\end{figure}

To further understand the relationship between internal aspects of LIPSAL when utilising GAI, a Pearson's correlation analysis was conducted. The results of the analysis are presented in Fig.\ref{correlation}. A significant positive correlation exists between all the variables, indicating that they are mutually reinforcing. The variable of learning interest demonstrated strong correlations with all the other five variables. This suggests that learning interest represents a central factor influencing not only learning performance but also the overall learning experience. Students with a high level of learning interest tend to exhibit a range of positive learning behaviours, including independent learning, problem solving, self-confidence, appropriate use  and learning enjoyment. A strong correlation was observed between learning interest and independent learning (r = 0.71, p \textless{} 0.01), as well as learning enjoyment (r = 0.76, p \textless{} 0.01). Additionally, a significant positive correlation was found between problem solving and learning enjoyment (r = 0.65, p \textless{} 0.01).  Then, we conducted an in-depth analysis on the two groups with high correlations (learning interest and independent learning, leaning interest and learning enjoyment) and the two groups with low correlations (self-confidence and problem solving, self-confidence and learning enjoyment), as illustrated in Fig.\ref{detail}. The score distribution of learning interest, independent learning and learning enjoyment is relatively concentrated, whereas self-confidence and problem solving are relatively dispersed. The distribution of scores is spread out and the correlation will be relatively weak. 

Overall, the results of the questionnaire indicate a general positive expectation of the potential for GAI to facilitate learning. The mean values of all variables were slightly above 4, indicating that participants generally rated the LIPSAL highly within the range investigated. It is evident that GAI has a considerable impact on the LIPSAL of student learning, although there are slight variations. This suggests that there are shortcomings in terms of fostering interest and self-confidence in the learning process. This correlation indicated that learning interest is conducive to a positive learning experience, which in turn facilitated student immersion and enhanced learning enjoyment. Furthermore, the utilisation of GAI by students to resolve the challenges they encounter enhanced the learning enjoyment. When viewed as a unified entity, the LIPSAL demonstrated a clear correlation, with each influencing student learning in a direct manner.  Therefore, it was imperative to continue examining the interrelationships between these variables, as they collectively facilitated the comprehensive growth and development of students.

\subsection{RQ2: What are the differences of LIPSAL between four grade levels in using GAI ?}
\label{4.2}
\begin{figure}[t]
\begin{tikzpicture}
\begin{axis}[
    ybar,
    enlargelimits=true,
    enlarge x limits=0.15,
    width = 14cm,
    height = 8cm,
    legend style={at={(0.5, -0.15)}, 
      legend columns=3,
      font=\small, % 设置图例字体大小
      anchor=north,
      legend cell align=left,
      % legend cell padding=2pt,
      },
    symbolic x coords={Sophomores in H.S, Juniors in H.S, Juniors in U.S, Seniors in U.S},
    xtick=data,
    bar width=11.5pt,
    nodes near coords,
    nodes near coords style={ % 设置柱子旁边数值的样式
        %font=\tiny, % 使用更小的字体
        scale = 0.65,
    },
    % nodes near coords align={vertical},
    % 添加虚线网格
    % grid = major,
    % grid style={dashed,gray!30},
    ]    
\addplot[pattern=horizontal lines] coordinates {(Sophomores in H.S,3.37) (Juniors in H.S,3.87) (Juniors in U.S,4.13) (Seniors in U.S,4.13)};
\addplot[pattern=crosshatch] coordinates {(Sophomores in H.S,3.4) (Juniors in H.S,3.93) (Juniors in U.S,4.33) (Seniors in U.S,4.23)};
\addplot[pattern=north east lines] coordinates {(Sophomores in H.S,3.4) (Juniors in H.S,3.56) (Juniors in U.S,4.19) (Seniors in U.S,4.3)};
\addplot[pattern=crosshatch dots] coordinates {(Sophomores in H.S,3.27) (Juniors in H.S,3.73) (Juniors in U.S,4.21) (Seniors in U.S,4.15)};
\addplot[pattern=vertical lines] coordinates {(Sophomores in H.S,3.69) (Juniors in H.S,4) (Juniors in U.S,4.2) (Seniors in U.S,4.22)};
\addplot[pattern=north west lines] coordinates {(Sophomores in H.S,3.42) (Juniors in H.S,3.76) (Juniors in U.S,4.17) (Seniors in U.S,4.27)};
\legend{learning interest, independent learning, problem solving, self-confidence,appropriate use, learning enjoyment}
\end{axis}
\end{tikzpicture}
\vspace*{0.2cm}
\centering 
\footnotesize
\text{Note: H.S represents High School and U.S represents University.} 
\caption{The mean value of LIPSAL across the four grades.}
\label{bar}
\end{figure}
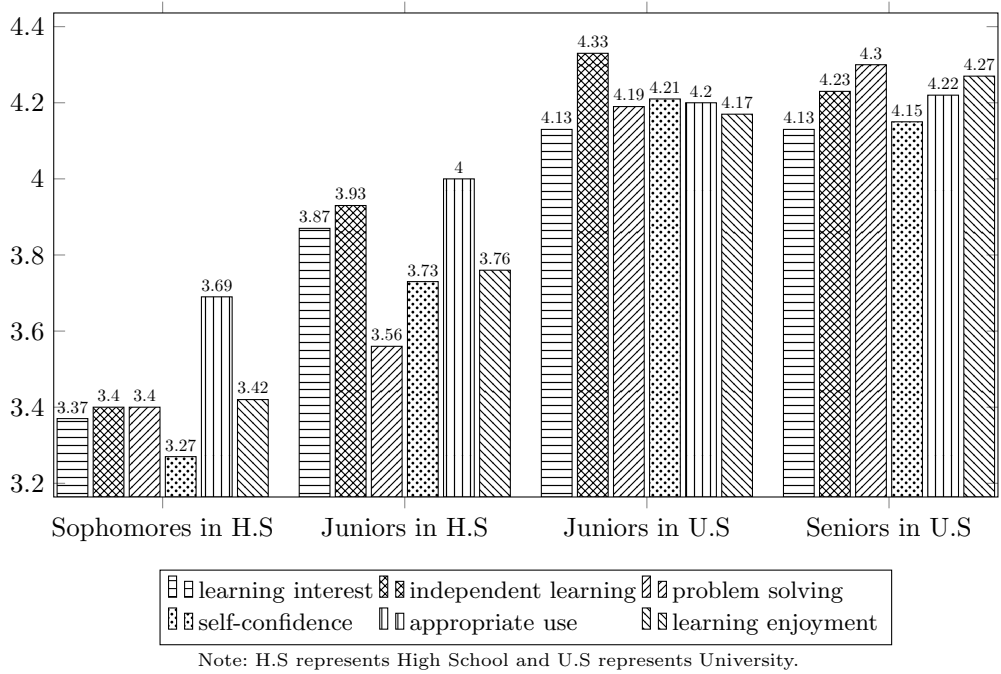

In order to address RQ2, we conducted the study involving sophomores in high school, juniors in high school, juniors in university and seniors in university. Fig.\ref{bar} illustrates the mean of the LIPSAL across the four grades. As evidenced by the graph, the overall means for university students were above 4 on the LIPSAL, whereas high school students exhibited generally lower means, below 4 on the LIPSAL. As students progress through their academic careers, they demonstrate growth in a multitude of areas. Such growth is particularly evident in the areas of self-confidence, independent learning and problem solving. 

The findings revealed that the mean for self-confidence (M=3.27) was the lowest, while the mean for appropriate use (M=3.69) was the highest, in the sophomore year of high school. It indicates that sophomore students in high school do not demonstrate an inclination towards engaging with problem-solving and independent learning within the GAI learning environment. Their attention is primarily directed towards the rational utilisation of GAI, with a continued emphasis on the learning application phase. In the senior year of high school, the mean for appropriate use (M=4) remained the highest, while the mean for problem solving (M=3.56) was found to be the lowest. This suggests that senior students in high school have not yet attained proficiency in utilising GAI for problem-solving purposes. However, compared to the preceding year, their capacity to learn independently is enhanced, thereby fostering greater self-confidence. There was a notable surge in all variables during the junior year in university. The mean for independent learning (M=4.33) was the highest, while the mean for interest in learning (M=4.13) was the lowest, although not significantly different from the other variables. It suggests that junior students in university achieve a more balanced level of development across all domains when utilising GAI, with a notable increase in independent learning abilities. In the fourth year of university, there is not a notable difference from the third year students, except that the mean value of problem solving reaches its maximum (M=4.3). This indicates that senior students in university are approaching graduation and are therefore focusing more on problem-solving skills,in order to better apply the theoretical knowledge they have acquired in practice. It is noteworthy that the mean for academic interest was the lowest of the LIPSAL, despite the fact that college students exhibited a higher level of interest overall than their high school counterparts. This indicates that college students do not prioritize learning interest when utilizing GAI. The typical college student is more inclined to engage in research and independent creation activities with well-defined objectives and superior problem-solving abilities. In comparison, they possess a certain degree of familiarity with and discernment regarding the role of GAI, so the value of learning interests is relatively minimal.

\begin{figure}[!ht]%% placement specifier
\centering%% For centre alignment of image.
\includegraphics[width=0.9\textwidth]{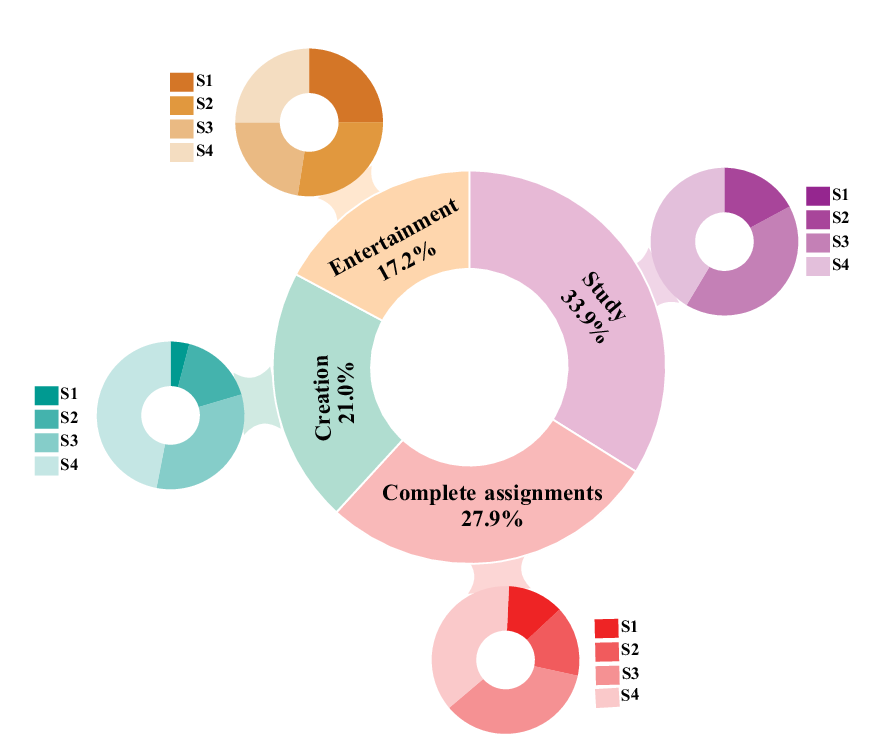}
% \caption{Percentage of GAI utilisation in four grades.}
\caption{Purposes of using GAI and proportion among students.}
\footnotesize
\noindent\begin{minipage}{\textwidth}
    \textbf{Note:} S1 represents Sophomores in H.S, S2 represents Juniors in H.S, S3 represents Juniors in U.S, S4 represents Seniors in U.S.
    \end{minipage}
\label{way}
\end{figure}

In order to gain insight into the differences in each variable across the four grades, a analysis of variance (ANOVA) was conducted. As demonstrated in Table.\ref{tab: mean}, a comparison of these data sets can reveal whether there are significant differences in the LIPSAL across grades. Learning interest (F=16.68, p=0.003), independent learning (F=10.67,  p=0.024), self-confidence (F=12.04, p=0.045) showed a significant differences. Through Fig.\ref{way}, we found that students utilise GAI more frequently for the purposes of studying and completing assignments. Furthermore, both of these activities are undertaken with greater frequency by juniors in university and seniors in university of study. This suggests that these students of study have a higher need for GAI than those  Sophomores in high school and juniors in high school, which in turn affects the level of development in the LIPSAL.

\begin{sidewaystable*}
\caption{ANOVA in the LIPSAL across grades.}
 \label{tab: mean}
    % \centering
    \small
    \footnotesize
\scalebox{1.1}
    {
    \setlength{\extrarowheight}{10pt}
    \begin{tabular}{l | c c | c c | c c | c c | c c }
    \toprule
    \multirow{2}{*}{\textbf{ LIPSAL}} & \multicolumn{2}{c}{\textbf{Sophomores in H.S}} & \multicolumn{2}{c}{\textbf{Juniors in H.S}} & \multicolumn{2}{c}{\textbf{Juniors in U.S}} & \multicolumn{2}{c}{\textbf{Seniors in U.S}} & \multirow{2}{*}{F}  & \multirow{2}{*}{Sig}
    \\
    \cmidrule(lr){2-3} \cmidrule(lr){4-5} \cmidrule(lr){6-7} \cmidrule(lr){8-9}
     & M & SD & M & SD & M & SD & M & SD & & \\
    \midrule
    \text{\textbf{Learning interest}} &3.37&0.35 &3.87&0.48&4.13&0.39&4.13&0.32&16.68&0.003 \\
    \text{\textbf{Independent learning}} &3.40&0.63 & 3.93 &0.59 &4.33&0.55&4.23&0.50&10.67&0.024 \\
    \text{\textbf{Problem solving}} &3.40&0.40 &3.56&0.47 &4.19&0.36&4.30&0.32&29.01&0.336 \\
    \text{\textbf{Self-confidence}} & 3.27&0.65&3.73&0.56& 4.22&0.49&4.15&0.56&12.04&0.045 \\
    \text{\textbf{Appropriate use}} & 3.69&0.43&4.00&0.44 &4.20&0.39&4.22&0.32&7.83&0.058 \\
    \text{\textbf{Learning enjoyment}} & 3.42&0.41&3.76&0.56 &4.17 &0.36&4.27&0.28&20.04&0.072\\
    \bottomrule
    \end{tabular}
    }

    \vspace{0.5em}
    % \centering 
    \footnotesize
    \text{\hspace{0.2cm} Note: M = mean value, SD = standard deviation, F = F-statistic, Sig = significance level.} 
\end{sidewaystable*}

In conclusion, the results demonstrated that university students exhibited a higher level in the LIPSAL compared to their secondary school counterparts,showing a more comprehensive and nuanced understanding of GAI. On the other hand,it highlights the importance of early training and guidance in the use of GAI, particularly in lower grades, to facilitate a more effective digital transformation in education. Proficiency in GAI usage from an early age is crucial for promoting effective learning. This helps to bridge the gap between the lower and upper classes and promotes GAI in learning.

\subsection{RQ3: What are the perceptions and understanding of students regarding GAI ?}
\label{4.3}
Guided by question 3, the study was divided into two parts. Firstly, an open-ended question was posed in the questionnaire, and the results were analysed for word frequency through the Easy Word Cloud \footnote{\url{https://www.yciyun.com/}}. Secondly, in-depth communication with students was conducted through group interviews, with the objective of obtaining more authentic and comprehensive results.

\begin{figure}[t]%% placement specifier
\centering%% For centre alignment of image.
\includegraphics[width=0.8\textwidth]{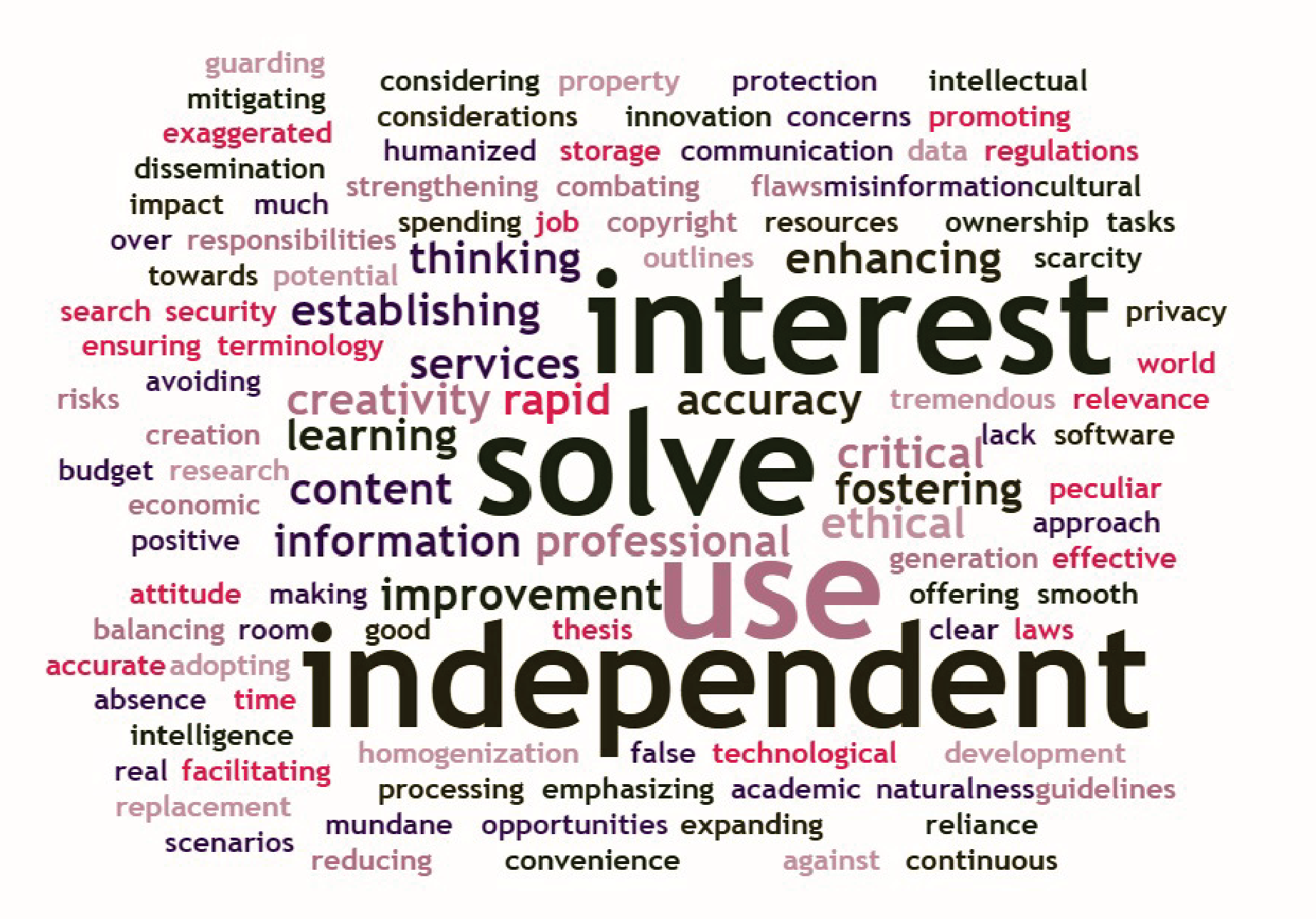}
\caption{Word cloud generated from the qualitative question.}
\label{Wordcloud}
\end{figure}

The results of the questionnaire (see Fig.\ref{Wordcloud}) indicate that the four most frequently occurring words are ‘\textit{interest}’, ‘\textit{use}’, ‘\textit{independent}’, and ‘\textit{solve}’. These words express students' positive attitudes and expectations towards GAI. For example, one student mentioned, ‘GAI provides content that is useful and interesting to me.’ Other students posited that the role of GAI in enhancing efficiency has facilitated the expeditious processing of voluminous academic data and associated information. Furthermore, the use of terms such as '\textit{potential}', '\textit{humanised}' and '\textit{considering}' suggests that students may hold a neutral stance. The student proposed, ‘To establish and improve relevant laws and regulations to regulate the development and use of GAI.’ It was also mentioned, ‘Hopefully it will be more humane and approachable.’ 

It is also evident that there are unsupported words such as '\textit{exaggerated}', '\textit{critical}' and '\textit{security}', which may serve to express students' concerns or negative perceptions. Students highlighted the potential risks associated with GAI, including the possibility of privacy breaches and the emergence of ethical concerns, discriminatory content, and other issues. It is worth noting that the terms ‘\textit{creativity}’, ‘\textit{rapid}’ and ‘\textit{content}’ that may be relevant to the application and development of GAI. The students indicated that they were able to rapidly generate original content and provide innovative inspiration. It has significant potential to facilitate solutions to complex challenges across a range of fields, including healthcare, education and research. By analysing these words we can see that they want to use GAI to solve problems and improve efficiency. However, this is accompanied by a concomitant concern regarding the potential adverse effects and constraints of GAI. It is therefore incumbent upon educators and those developing GAI to collaborate in order to guarantee that it is deployed in a secure, ethical and holistic manner.

\begin{sidewaystable*}
\caption{Students' understanding and perceptions of GAI.}
 \label{tab: perspectives}
    % \centering
    % \small
    % \footnotesize
\scalebox{1.1}
    {
    \setlength{\extrarowheight}{3pt}
    \begin{tabular}{p{6cm}@{\hspace{1cm}} p{10cm}}
    \toprule
\textbf{Learning experience} \\
What are the positive impacts when you use GAI? & $\bullet$ Increases efficiency and convenience \\
& $\bullet$ Helps us solve problems better and think independently \\
& $\bullet$ Can be connected to more specialised knowledge to provide better service \\
& $\bullet$ Helps us understand the great world and learn interesting content \\
What difficulties and challenges have you encountered when you use GAI? & $\bullet$ Inaccurate feedback \\
& $\bullet$ Not smart and user-friendly enough \\
& $\bullet$ It can lead some people to become overly dependent, thus weakening their own ability to think and create \\
& $\bullet$ Exist intellectual property issues such as plagiarism \\
\textbf{Perceptions} \\
What is your attitude towards using GI? & $\bullet$ Generally positive attitude \\
What are your thoughts on the future of GAI? & $\bullet$ Schools need to develop critical thinking in their students and make reasonable use of careful understanding of the GAI \\
& $\bullet$ Enhanced technological development to improve content accuracy and reduce errors\\
& $\bullet$ Before using a GAI tool, users should be clear about their needs and goals, which will help the tool generate the desired content more accurately \\
    \bottomrule
    \end{tabular}
    } 
\end{sidewaystable*}

In order to gain a deeper insight into the influence of GAI on students, further research and analysis is required. We randomly selected 20 students to be interviewed about the positive impacts, difficult challenges, their own attitudes and the future development of GAI. The results are displayed in Table.\ref{tab: perspectives},  the findings of the interviews indicate that students hold a generally positive attitude towards GAI. They perceive that GAI enhances efficiency, facilitates learning, and provides solutions to problems. Additionally, the students indicated that GAI might assist them in comprehending a more extensive array of content. However, when utilising GAI, students identified that the system provided inaccurate information and feedback, and was not yet sufficiently intelligent or user-friendly. There is another concern that GAI may lead to intellectual property issues such as plagiarism. It may result in an over-reliance on GAI and a subsequent reduction in the capacity of individuals to think and create independently. These interviewees made several suggestions, 1) schools should endeavour to foster critical thinking in their students and make prudent use of GAI, 2) technological advances could enhance the accuracy of GAI's content and reduce errors, 3) users define their requirements and objectives prior to utilising GAI tools, in order to facilitate the more precise generation of the requisite content.

Overall, the students demonstrated a favourable attitude and a comprehensive understanding of the application of GAI. However, certain matters remained requiring further attention. This necessitates collaboration between educational establishments and society to provide guidance and training to students on the appropriate utilisation of GAI as a whole. Use it to provide personalised instruction, corrective feedback and guidance when students encounter difficulties, and the cultivation of independent learning abilities. Furthermore, a variety of engaging knowledge can be presented to enhance students' learning interest and enjoyment. Additionally, efforts must be made to enhance digital literacy and critical thinking skills in students, thus enabling a more effective adaptation to future societal developments.

\section{Conclusion}\label{sec5}

In this paper, a hybrid-survey research method was proposed to gain insight into the impact of GAI on students and their perceptions of this new technology. By analysing the results of questionnaires and interviews, we found that the GAI-assisted learning facilitates the development of the LIPSAL among all students. Following are the main specific findings. Firstly, college students performed better than middle school students on the LIPSAL, indicating a better understanding of the GAI. This also demonstrates the need for training and guidance in the use of GAI in lower grades, in order to encourage better transformation of digital in education. Secondly, students had a positive willingness to utilise GAI. This could be the reason for GAI's rapid expansion in education. However, it also revealed that students have difficulty using GAI in an appropriate way. They are also less proficient in utilising this technological tool to enhance learning and optimise learning outcomes. There is a need for teachers and schools to teach students how to use GAI properly and develop independent learning skills. Thirdly, GAI itself has some shortcomings due to it's only been out about 2 years, such as lack of intelligence and inaccurate results. This requires continuous improvement by technologists based on user feedback, so that it can be deeply integrated into education and drive the digital transformation of education.

\textbf{Limitations:} It should be noted that this study is subject to certain limitations. On the one hand, the study was conducted with Beijing students in an online setting. As a result, it is not readily applicable to other academic levels. On the other hand, the sample size was relatively small and homogeneous. Future research should aim to address these limitations by employing a larger and more diverse sample, including students from various regions and academic levels. The incorporation of teachers and schools into the study population would also serve to enhance the quality of the data. Furthermore, it should provide a more comprehensive understanding of the factors influencing students' educational experiences.

\bibliography{References}% common bib file

\section*{Competing interests}
The authors declare no competing interests.
\section*{Data availability}
The research data associated with this paper can be accessed at \url{https://drive.google.com/drive/folders/10pIZtCaOZMAJBiqoiOy635y00-PvMUzI?usp=drive_link}.
\section*{Ethical approval}
All procedures performed in this study were in accordance with the ethical standards stipulated in the Declaration of Helsinki and its subsequent amendments. Ethical clearance and approval were granted by School of Mathematics and Statistics, Guangxi Normal University, on Aug 11th, 2024.
\section*{Informed consent}
Informed consent for this study was obtained via the Wenjuanxing platform before data collection began on Aug 11th, 2024. Participants were provided with detailed information about the study and asked to give their consent by selecting the appropriate option in the online questionnaire. They were fully informed about the study’s nature and purpose, with assurance that the data would be used exclusively for academic purposes. Participant anonymity is guaranteed, as no personally identifiable information is collected or stored, and their responses would remain confidential. Additionally, participants were notified that there are no foreseeable risks involved in the study and that they could withdraw at any point without facing any consequences. 
% \section*{Author contributions}

\end{document}